\documentstyle[prl,aps,twocolumn]{revtex}
\begin{document}
\draft
\title{Many-Body Theory of Dilute Bose-Einstein Condensates 
with Internal Degrees of Freedom}
\author{Masahito Ueda}
\address{Department of Physics,
Tokyo Institute of Technology,
2-12-1 Ookayama, Meguro-ku, Tokyo 152-8551, Japan}
\date{\today}
\maketitle
\begin{abstract}
The Bogoliubov theory of weakly interacting bosons is generalized to 
Bose-Einstein condensates with internal degrees of freedom so that a 
single effective Hamiltonian produces various many-body ground states
or metastable spin domains and the corresponding collective modes on 
an equal footing.
\end{abstract}

\narrowtext

The internal degrees of freedom of alkali Bose-Einstein condensates (BECs)
arise primarily from electronic spin and are hence by far more amenable to
manipulation than those of superfluid $^3$He. 
The realization of ^^ ^^ spinor" BECs~\cite{Stenger} has motivated many  
researchers to study the magnetism of superfluid vapors
~\cite{Ohmi,Ho,Law,Koashi,HoYip,Isoshima}.
The excitation spectra of spinor BECs have been examined by Ohmi and 
Machida~\cite{Ohmi} and by Ho~\cite{Ho} using mean-field theory.  
The many-body ground states were investigated by Law {\it et al.}~\cite{Law}
and by Koashi and Ueda~\cite{Koashi} and Ho and Yip~\cite{HoYip} in the absence
and presence, respectively, of a magnetic field; in the latter case, it is
predicted that the magnetic sublevel $m=0$ of a spin-1 antiferromagnetic BEC 
becomes populated as the magnetic field decreases. 
The main aim of this Letter is to present a unified theory that derives the
various many-body states of BEC and the corresponding excitation spectra on
an equal footing.
Previous work~\cite{Law,Koashi,HoYip} has been unable to address this problem,
as it has been assumed that the internal degrees of freedom are independent of
other degrees of freedom such as the density and spin waves. 

We consider a system of $N$ spin-1 identical bosons interacting via a binary, 
contact interaction, subject to a uniform magnetic field ${\bf B}$. 
The Hamiltonian of the system is given by
\begin{eqnarray}
\hat{H}&=&\int d{\bf r}\left[\hat{\Psi}_\alpha^\dagger\left(\!
-\frac{\hbar^2}{2M}\nabla^2\!+U({\bf r})\!\!\right) \hat{\Psi}_\alpha
-p\hat{\Psi}_\alpha^\dagger f^z_{\alpha\beta}
\hat{\Psi}_\beta
\right. \nonumber\\
& & \left.
+\frac{c_0}{2}\hat{\Psi}_\alpha^\dagger\hat{\Psi}_\beta^\dagger
\hat{\Psi}_\beta\hat{\Psi}_\alpha+
\frac{c_1}{2}\hat{\Psi}_\alpha^\dagger\hat{\Psi}_{\alpha'}^\dagger
{\bf f}_{\alpha\beta}\cdot{\bf f}_{\alpha'\beta'}
\hat{\Psi}_{\beta'}\hat{\Psi}_\beta
\right],
\label{H}
\end{eqnarray}
where $p(>0)$ is the product of the gyromagnetic ratio 
and the external magnetic field $B$ applied in the z-direction;
${\bf f}_{\alpha\beta}=(f_{\alpha\beta}^x,f_{\alpha\beta}^y,f_{\alpha\beta}^z)$
represents the components ($\alpha,\beta=1,0,-1$) of spin-1 matrices; 
and $\hat{\Psi}_\alpha({\bf r})$ denotes the field operator that annihilates 
a boson in magnetic sublevel $\alpha$ at position ${\bf r}$.
The quadratic Zeeman effect will be taken into account when it becomes 
relevant.
The coupling constants $c_0$ and $c_1$ are related to the spin-singlet and 
triplet s-wave scattering lengths $a_s$ and $a_t$ by
$c_0=4\pi\hbar^2(a_s+2a_t)/3M$ and $c_1=4\pi\hbar^2(a_t-a_s)/3M$, respectively.
In Eq.~(\ref{H}) the repeated indicies are assumed to be summed.

When the system has translation invariance, it is convenient to expand 
the field operator as 
$\hat{\Psi}_\alpha({\bf r})=(1/\sqrt{V})\sum_{\bf k}
\hat{a}_{{\bf k}\alpha}e^{i{\bf kr}}$,
where $V$ is the volume of the system.
Equation~(\ref{H}) then reduces to
\begin{eqnarray}
\hat{H}=\!\sum_{{\bf k}\alpha}(\epsilon_{\bf k}\!-\!p\alpha)
\hat{n}_{{\bf k}\alpha}
\!+\frac{1}{2V}\!\sum_{\bf k}:\!(
  c_0\hat{\rho}_{\bf k}^\dagger\hat{\rho}_{\bf k}
\!+\! c_1\hat{\bf S}_{\bf k}^\dagger\! \cdot\!\hat{\bf S}_{\bf k})
\!:,
\label{H2}
\end{eqnarray}
where 
$\epsilon_{\bf k}\equiv\hbar^2{\bf k}^2/2M$, 
$\hat{n}_{{\bf k}\alpha}\equiv
\hat{a}_{{\bf k}\alpha}^\dagger\hat{a}_{{\bf k}\alpha}$,
$\hat{\rho}_{\bf k}\equiv\sum_{\bf q}\hat{a}_{{\bf q}\alpha}^\dagger
\hat{a}_{{\bf q+k}\alpha}$,
$\hat{\bf S}_{\bf k}\equiv\sum_{{\bf q}\alpha\beta}{\bf f}_{\alpha\beta}
\hat{a}_{{\bf q}\alpha}^\dagger\hat{a}_{{\bf q+k}\beta}$,
and $::$ denotes normal ordering.

When the interparticle interactions are repulsive, BEC occurs in the 
${\bf k}=0$ state. We therefore separate the ${\bf k}=0$ components  
$\hat{a}_{{\bf 0}\alpha}$ in the Hamiltonian and keep the terms only 
up to second order in $\hat{a}_{{\bf k}\alpha}$ (${\bf k}\neq 0$).
Because our interest is to derive various many-body states and the
corresponding low-lying excitation spectra on an equal footing, we 
should not replace the operators  $\hat{a}_{{\bf 0}\alpha}$ 
by c-numbers.
After some algebraic manipulation, we obtain~\cite{Ueda}
\begin{eqnarray}
& & \hat{H}^{\rm eff}=\frac{c_0+c_1}{2V}N(N-1)
-\frac{3c_1}{2V}\hat{A}^\dagger\hat{A}
-p(\hat{n}_{{\bf 0}1}-\hat{n}_{{\bf 0}-1}) 
\nonumber \\
& & 
+\sum_{\bf k\neq0}\sum_{\alpha=0,\pm1}\!\left(
\epsilon_{\bf k}-p\alpha-\frac{(1+|\alpha|)c_1}{V}
\hat{n}_{{\bf 0}-\alpha}\right)
\hat{n}_{{\bf k}\alpha}
\nonumber \\
& & 
+\frac{1}{V}\sum_{\bf k\neq0}\left\{\frac{c_0}{2}
(\hat{D}_{\bf k}\hat{D}_{\bf -k}
+\hat{D}_{\bf k}^\dagger\hat{D}_{\bf k})
+\frac{c_1}{2}(\hat{S}_{\bf k}\hat{S}_{\bf -k}
+\hat{S}_{\bf k}^\dagger\hat{S}_{\bf k})\right.
\nonumber \\
& & 
+\sum_{\alpha=0,1}\!c_1[
\hat{a}_{{\bf k}\alpha}^\dagger\hat{a}_{{\bf -k}-\alpha}^\dagger
\hat{a}_{{\bf 0}1-\alpha} \hat{a}_{{\bf 0}\alpha-1}
+ \hat{a}_{{\bf k}\alpha-1}^\dagger\hat{a}_{{\bf -k}\alpha}^\dagger
\hat{a}_{{\bf 0}\alpha-1}\hat{a}_{{\bf 0}\alpha}
\nonumber \\
& & \left.
+ (\hat{a}_{{\bf k}\alpha}^\dagger\hat{a}_{{\bf k}\alpha-1}
+2\hat{a}_{{\bf k}1-\alpha}^\dagger\hat{a}_{{\bf k}-\alpha})
\hat{a}_{{\bf 0}\alpha-1}^\dagger\hat{a}_{{\bf 0}\alpha}
]+{\rm h.c.}\right\}
\label{Bog}
\end{eqnarray}
where h.c. denotes the hermitian conjugates of all preceding terms in the
curly brackets,
$\hat{A}^\dagger\equiv(\hat{a}_{{\bf 0}0}^{2\dagger}-2\hat{a}_{{\bf 0}1}^\dagger\hat{a}_{{\bf 0}-1}^\dagger)/\sqrt{3}$
creates a pair of spin-singlet bosons when acting upon the vacuum, and
$\hat{D}_{\bf k}\equiv\sum_\alpha\hat{a}_{0\alpha}^\dagger\hat{a}_{{\bf k}\alpha}$ and 
$\hat{S}_{\bf k}\equiv\sum_\alpha\alpha\hat{a}_{0\alpha}^\dagger\hat{a}_{{\bf k}\alpha}$ describe density-wave and spin-wave operators, respectively.
The first three terms on the right-hand side of Eq.~(\ref{Bog}) were 
studied in Refs.~\cite{Law,Koashi,HoYip}; the remaining terms show 
how the internal and spatial degrees of freedom couple to one another.

To set a reference frame for later discussions, let us recapitulate the
main results of Refs.\cite{Law,Koashi,HoYip}. The Hamiltonian discussed
in these references reads essentially 
\begin{eqnarray}
& & \hat{H}_0=
-\frac{3c_1}{2V}\hat{A}^\dagger\hat{A}
-p(\hat{n}_{{\bf 0}1}-\hat{n}_{{\bf 0}-1}).
\label{eff}
\end{eqnarray}
When $c_1<0$, the energy minimum is obtained by putting all particles in 
the state with ${\bf k}={\bf 0}$ and $\alpha=1$:
\begin{eqnarray}
|N_2=0,F=N,F_z=N\rangle
=\frac{(\hat{a}_{{\bf 0}1}^\dagger)^N}{\sqrt{N!}}
|{\rm vac}\rangle,
\label{f}
\end{eqnarray}
where $F$ and $F_z$ denote the total spin and its projection on the 
$z$ axis, and $N_2$ is given in terms of $F$ and the total number of
atoms $N$ by $N_2=(N-F)/2$.
Because all spins are aligned in the same direction, the system is 
ferromagnetic.
When $c_1>0$, the first term on the right-hand side of Eq.~(\ref{eff}) 
energetically favors the spin-singlet (or antiferromagnetic) correlation, 
whereas the second one favors the parallel-spin configurations. 
For a given external magnetic field ($\propto p$), 
the energy minimum is attained for the state~\cite{Koashi}
\begin{eqnarray}
& & |N_2=(N-F)/2,F,F_z=F\rangle_0 \nonumber \\
&\propto&(\hat{A}^\dagger)^{N_2}(\hat{F}_{-})^{F-F_z}
(\hat{a}_{{\bf 0}1}^\dagger)^F|{\rm vac}\rangle,
\label{af}
\end{eqnarray}
where $F$ denotes an integer that satisfies
$F-1/2<p/c_1<F+3/2$, and
$\hat{F}_{-}\equiv(f^x-if^y)_{\alpha\beta}\hat{a}_{{\bf 0}\alpha}^\dagger
\hat{a}_{{\bf 0}\beta}$. 
In both Eqs.~(\ref{f}) and (\ref{af}), the depletion of the condensate due
to interactions is not taken into account.

When the components with ${\bf k}={\bf 0}$ are macroscopically occupied, 
we may replace $\hat{a}_{{\bf 0}\alpha}$ in Eq.~(\ref{Bog}) with the 
c-numbers $\sqrt{N}\zeta_\alpha$. 
To find the correct excitation spectra, it is crucial to take the 
depletion of the condensate into account.
This is done by setting
\begin{eqnarray}
\sum_\alpha|\zeta_\alpha|^2=1-\frac{1}{N}\sum_{{\bf k}\neq{\bf 0},\alpha}\hat{n}_{{\bf k}\alpha},
\label{deplete}
\end{eqnarray}
where the last term describes the fraction of the depletion.
The energy of the condensate is given by
\begin{eqnarray}
E_0=-\frac{c_1N^2}{2V}|\zeta_0^2-2\zeta_1\zeta_{-1}|^2-pN(|\zeta_1|^2
-|\zeta_{-1}|^2),
\label{E0}
\end{eqnarray}
where 
$\zeta_\alpha$ are determined by requiring that $E_0$ be minimized 
subject to constraint (\ref{deplete}).

For $c_1<0$, the minimum of $E_0$ is 
reached when
\begin{eqnarray}
\zeta_0=\zeta_{-1}=0, \ \ 
|\zeta_1|^2=1-\frac{1}{N}
\sum_{{\bf k}\neq{\bf 0},\alpha}\hat{n}_{{\bf k}\alpha}.
\end{eqnarray}
The Hamiltonian (\ref{Bog}) then reduces to
\begin{eqnarray}
& & \hat{H}^{\rm F}=
\frac{c_0+c_1}{2V}N(N-1)-pN \nonumber \\
& & +
\sum_{{\bf k}\neq0}
\left[\epsilon_{\bf k}\hat{n}_{{\bf k}1}
+\frac{(c_0+c_1)N}{2V}
(\hat{a}_{{\bf k}1}\hat{a}_{{\bf -k}1}+\hat{n}_{{\bf k}1}+{\rm h.c.})
\right.\nonumber \\
& & \left.
+ (\epsilon+p)\hat{n}_{{\bf k}0}
+ (\epsilon+2p-\frac{2c_1N}{V})\hat{n}_{{\bf k}-1}\right].
\label{F}
\end{eqnarray}
Diagonalizing this Hamiltonian, we find the excitation spectra as
\begin{eqnarray}
E_{{\bf k},1}^{\rm F} &=&
\sqrt{\epsilon_{\rm k}[\epsilon_{\rm k}+2(c_0+c_1)n]}, \nonumber \\
E_{{\bf k},0}^{\rm F} &=&\epsilon_{\bf k}+p,       \nonumber \\
E_{{\bf k},-1}^{\rm F}&=&\epsilon_{\bf k}+2p-2c_1n,
\end{eqnarray}
in agreement with Ref.\cite{Ohmi}. 
We may gain some insight into the nature of the many-body correlation by
writing the many-body wave function in the coordinate representation:
\begin{eqnarray}
\Psi({\bf r}_1\alpha_1,\cdots,{\bf r}_N\alpha_N)=\langle{\rm vac}|
\hat{\Psi}_{\alpha_1}({\bf r}_1)\cdots\hat{\Psi}_{\alpha_1}({\bf r}_1)
|\Phi\rangle.
\end{eqnarray}
Here $|\Phi\rangle$ is given in the Bogoliubov approximation by
\begin{eqnarray}
|\Phi\rangle\sim\exp\left(\phi_0\hat{a}_{{\bf 0}1}^\dagger-
\sum_{k_x>0}\nu_k\hat{a}_{{\bf k}1}^\dagger
\hat{a}_{-{\bf k}1}^\dagger\right)
|{\rm vac}\rangle,
\end{eqnarray}
where $\phi_0^2=N[1-(8/3)\sqrt{na_{\rm t}^3/\pi}]$ and 
$\nu_k=1+c_k-\sqrt{c_k(c_k+2)}$ with $c_k\equiv k^2/(8\pi a_{\rm t}n)$.
In the dilute limit, the many-body wave function becomes very small unless
all $\alpha_k$'s are equal 1, and we find that~\cite{Ueda}
\begin{eqnarray}
& & \Psi({\bf r}_11,\cdots,{\bf r}_N1)\simeq 
\frac{\exp\!\left(-\frac{4}{9}(3\pi-8)N\sqrt{na_t^3/\pi}
\right)}{(2\pi NV^{2N})^{1/4}} \nonumber \\
& & \times
\exp\left(-\sum_{i<j}\frac{a_t}{r_{ij}}
e^{-r_{ij}/\xi}
\right),
\label{mbwf}
\end{eqnarray}
where $r_{ij}\equiv|{\bf r}_i-{\bf r}_j|$ and the terms that are of the order
of $1/N$ are ignored.
This result clearly shows that two bosons strongly repel each other when
their distance becomes smaller than the healing length 
$\xi\equiv(8\pi a_tn)^{-1/2}$.

For $c_1>0$, minimizing $F=E_0-\delta\sum_\alpha|\zeta_\alpha|^2$, 
where $\delta$ is a Lagrange multiplier, yields
\begin{eqnarray}
|\zeta_0|\left(e^{i(\phi_1+\phi_{-1}-2\phi_0)}
+\frac{\delta}{\sqrt{\delta^2-\gamma^2}}\right)=0,
\label{afcond}
\end{eqnarray}
where $\phi_\alpha\equiv{\rm arg}\zeta_\alpha$ and $\gamma\equiv 2p/(c_1n)$.
In the presence of the external magnetic field (i.e., $\gamma\neq0$), 
the minimum of $F$ is attained when
\begin{eqnarray}
\zeta_0=0, \ \ 
|\zeta_{\pm1}|^2=\frac{1}{2}\pm\frac{\gamma}{4}-\frac{1}{2N}
\sum_{{\bf k}\neq0\alpha}\hat{n}_{{\bf k},\alpha},
\end{eqnarray}
and the corresponding Hamiltonian reads
\begin{eqnarray}
& & \hat{H}^{\rm AF}_{\gamma\neq0}=
\frac{c_0}{2V}N(N-1)+\frac{c_1}{2V}N\left(\frac{\gamma^2N}{4}-1\right)
-\frac{p\gamma N}{2} \nonumber \\
& & \sum_{{\bf k}\neq{\bf 0},\alpha}
(\epsilon_{\bf k}+c_1n\delta_{\alpha,0})\hat{n}_{{\bf k}\alpha}
+\frac{1}{2V}\sum_{{\bf k}\neq0}\left\{
c_0(\hat{D}_{\bf k}\hat{D}_{\bf -k}
+\hat{D}_{\bf k}^\dagger\hat{D}_{\bf k})
\right.\nonumber \\
& & \left.
+c_1(\hat{S}_{\bf k}\hat{S}_{\bf -k}
+\hat{S}_{\bf k}^\dagger\hat{S}_{\bf k}
+n|\zeta_1\zeta_{-1}|\hat{a}_{{\bf k},0}\hat{a}_{{\bf -k},0}
)
+ {\rm h.c.}\right\}.
\label{AF1}
\end{eqnarray}
The $\alpha=0$ mode is decoupled and its dispersion relation is given by
\begin{eqnarray}
E_{{\bf k},0}^{\rm AF}=\sqrt{
\epsilon_{\rm k}^2+2c_1n\epsilon_{\rm k}+p^2}.
\label{af0}
\end{eqnarray}
The $\alpha=\pm1$ modes are coupled, and the dispersion relations of the
coupled modes are given by
\begin{eqnarray}
E_{{\bf k},\pm}^{\rm AF}=\!\sqrt{
\epsilon_{\rm k}^2\!+2(c_0\!+\!c_1)n\epsilon_{\rm k}\!
\pm\epsilon_{\rm k}\sqrt{n^2(c_0\!-\!c_1)^2\!\!+\!\frac{4p^2c_0}{c_1}}},
\nonumber\\
\label{af1}
\end{eqnarray}
again in agreement with Ref.~\cite{Ohmi}.
Our theory thus reproduces all known collective modes 
that are derived using a different method.

An interesting situation arises in the absence of a magnetic field 
(i.e., $\gamma=0$), 
where condition (\ref{afcond}) can be met if 
\begin{eqnarray}
\phi_1+\phi_{-1}-2\phi_0=\pi.
\label{coh}
\end{eqnarray}
This result is in accordance with a rule pointed out in Ref.~\cite{Ueda2}, 
that is, when the interaction is attractive, the relative phase coherence,
as implied by the constraint (\ref{coh}), will spontaneously emerge if more 
than two BECs coexist. 
In the present case, the attractive force applies among three spin 
components, as implied by the condition $c_1>0$ (see Eq.~(\ref{eff})). 

When the relation (\ref{coh}) holds, Eq.~(\ref{afcond}) implies that 
$|\zeta_0|$ is arbitrary, so we have $|\zeta_1|=|\zeta_{-1}|,$ 
$|\zeta_0|^2=1-2|\zeta_1|^2$.
This, combined with Eq.~(\ref{coh}), leads to a vectorial order parameter as
${\bf \zeta}=(\zeta_1,\zeta_0,\zeta_{-1})
=e^{i\phi_0}(-e^{i(\phi_0-\phi_{-1})}\sin\beta/\sqrt{2},\cos\beta,
e^{-i(\phi_0-\phi_{-1})}\sin\beta/\sqrt{2})$. 
This result agrees with Eq.~(5) of Ref.~\cite{Ho} and implies that the spin 
components can change under spatial rotation. 
It is surprising that the many-body ground state predicts otherwise:
from Eq.~(\ref{af}), we find that at zero magnetic field (hence $F=0$) 
spin components cannot change under spatial rotation but must always be 
the same
~\cite{Koashi,HoYip} and is given by
$\hat{n}_{{\bf 0}\alpha}
=\left(N-\sum'_{{\bf k}\alpha}\hat{n}_{{\bf k}\alpha}\right)/3$ for 
$\alpha=0,\pm 1$,
where we take into account the depletion of the condensate 
$\sum'_{{\bf k}\alpha}\hat{n}_{{\bf k}\alpha}$ due to interactions.
This is a consequence of the fact that the true ground state is composed
of spin-singlet ^^ ^^ pairs" and is therefore invariant under rotation.
The above result for $\hat{n}_{{\bf 0}\alpha}$
 and the relative phase relation (\ref{coh}) may be used to 
drastically simplify the Hamiltonian (\ref{Bog}), giving
\begin{eqnarray}
& & \hat{H}^{\rm AF}_{\gamma=0}=
\frac{c_0}{2V}N(N-1)-\frac{c_1N}{2V}
\nonumber\\
& & 
+\sum_{{\bf k}\neq0}\left[ 
(\epsilon_{\bf k}\!+c_0n)\hat{d}_{{\bf k}}^\dagger\hat{d}_{\bf k}
+ (\epsilon_{\bf k}+c_1n)(\hat{s}_{{\bf k}}^\dagger\hat{s}_{\bf k}
+ \hat{q}_{{\bf k}}^\dagger\hat{q}_{\bf k})
\right.
\nonumber \\
& & \left.
+\frac{n}{2}\!(c_0\hat{d}_{\bf k}\hat{d}_{\bf -k}
+c_1\hat{s}_{\bf k}\hat{s}_{\bf -k}
+c_1\hat{q}_{\bf k}\hat{q}_{\bf -k}+{\rm h.c.})
\right],
\label{AF2}
\end{eqnarray}
where
\begin{eqnarray}
\hat{d}_{\bf k}&=& \frac{1}{\sqrt{3}} 
(\hat{a}_{{\bf k}1}e^{-i\phi_1}+\hat{a}_{{\bf k}0}e^{-i\phi_0}+\hat{a}_{{\bf k}-1}e^{-i\phi_{-1}}),
\nonumber \\
\hat{s}_{\bf k}&=& \frac{1}{\sqrt{2}}(\hat{a}_{{\bf k}1}e^{-i\phi_1}-\hat{a}_{{\bf k}-1}e^{-i\phi_{-1}}),
\nonumber \\
\hat{q}_{\bf k}&=& \frac{1}{\sqrt{6}}(\hat{a}_{{\bf k}1}e^{-i\phi_1}-2\hat{a}_{{\bf k}0}e^{-i\phi_0}+\hat{a}_{{\bf k}-1}e^{-i\phi_{-1}}).
\label{dsq}
\end{eqnarray}
We thus find that the density, spin, and quadrupolar fluctuations provide 
independent excitations.
The novelty of these modes is that they are phase-locked to the spin 
components of the condensate.
Diagonalizing the Hamiltonian (\ref{AF2}), we obtain the following excitation
spectra:
\begin{eqnarray}
E_{d,{\bf k}}^{\rm AF} &=&
\sqrt{\epsilon_{\rm k}[\epsilon_{\rm k}+2c_0n]}, \nonumber \\
E_{s,{\bf k}}^{\rm AF} &=&
E_{q,{\bf k}}^{\rm AF}  =
\sqrt{\epsilon_{\rm k}[\epsilon_{\rm k}+2c_1n]}.
\end{eqnarray}
In the thermodynamic limit, these collective modes survive only
at exactly zero magnetic field, and may be viewed as singular. 
However, in mesoscopic situations, e.g., when the system is 
confined in a quasi-one-dimensional torus (to which the present theory
applies with minor modifications), these modes survive at small 
magnetic fields, provided the ${\bf k}={\bf 0}$, $\alpha=0$ mode 
is macroscopically occupied (for a more precise 
definition of ^^ ^^ small" magnetic fields, see Ref.~\cite{Koashi}).

Recently, an MIT group has observed the formation of metastable spin 
domains~\cite{Miesner}.
They first prepared all atoms in the $\alpha=1$ state and then placed half
of them in the $\alpha=0$ state by irradiating the rf field. Letting 
the system evolve freely while using the quadratic Zeeman effect to  
prevent the $\alpha=-1$ component from appearing, they found that spin domains
formed with the two components alternatively aligned. 
It was discussed~\cite{Miesner,Meystre,Mueller} that this phenomenon 
is due to the imaginary frequencies of the excitation modes. Here we 
use our theory to confirm this hypothesis and to derive a general expression
for the dispersion relation.
The experimental conditions in effect amount to setting
\begin{eqnarray}
|\zeta_1|^2=|\zeta_0|^2=\frac{1}{2}-\frac{1}{2N}\sum_{{\bf k}\neq{\bf 0}}
(\hat{n}_{{\bf k}1}+\hat{n}_{{\bf k}0}), \ \ \zeta_{-1}=0.
\end{eqnarray}
Our Hamiltonian (\ref{Bog}) reduces to 
\begin{eqnarray}
& & \hat{H}=\frac{c_0+c_1}{2}n(N-1)-\frac{c_1}{8}nN
\nonumber \\
& & +\sum_{{\bf k}\neq{\bf 0}}\left[
\left(\epsilon_{\bf k}+\frac{c_0n}{2}+\frac{3c_1n}{4}\right)\hat{n}_{{\bf k}1}
+\left(\epsilon_{\bf k}+\frac{c_0n}{2}-\frac{c_1n}{4}\right)\hat{n}_{{\bf k}1}
\right.\nonumber \\
& & 
+\frac{c_0}{8}n(\hat{a}_{{\bf k}0}\hat{a}_{{\bf -k}0}+{\rm h.c.})
\nonumber \\
& & \left.
+\frac{c_0+c_1}{8}n(\hat{a}_{{\bf k}1}\hat{a}_{{\bf -k}1}
+\hat{a}_{{\bf k}1}\hat{a}_{{\bf -k}0}
+\hat{a}_{{\bf k}1}^\dagger\hat{a}_{{\bf -k}0}
+{\rm h.c.})
\right].
\end{eqnarray}
Diagonalizing the Hamiltonian, we find the dispersion relations as
\begin{eqnarray}
(E_{\bf k})^2&=&\epsilon_{\bf k}^2+(2u+v)\epsilon_{\bf k}
+\frac{3}{4}v^2 \pm\left[4(u^2+2uv+2v^2)\epsilon_{\bf k}^2
\right.\nonumber\\
& & 
\left.
+2v^2(2u+v)\epsilon_{\bf k}-v^3\left(u+\frac{3}{4}v\right)
\right]^\frac{1}{2}, 
\label{meta} 
\end{eqnarray}
where $u\equiv c_0n/2$ and $v\equiv c_1n/2$.
For parameters of the MIT experiment $n\sim10^{14}$ cm$^{-3}$, 
$a_{\rm t}\sim29$\AA, and $a_{\rm s}\sim26$\AA~\cite{Burke},
we find that $v/u\ll 1$. Ignoring in Eq.~(\ref{meta})
higher-order powers of $v/u$, we obtain
\begin{eqnarray*}
(E_{\bf k})^2=\epsilon_{\bf k}^2+(2u+v)\epsilon_{\bf k}\pm
2(u+v)\epsilon_{\bf k},
\end{eqnarray*}
where the plus sign
corresponds to the density wave,
while the minus sign corresponds to the spin wave.
We see that the energy of the spin wave becomes pure imaginary for 
$\epsilon_{\bf k}<v$, implying the formation of spin domains.  
The corresponding wavelength 
defines the characteristic length scale of the spin domains as
\begin{eqnarray}
\lambda_{\rm c}=\sqrt{\frac{3\pi}{(a_t-a_s)n}}.
\end{eqnarray}
This result agrees with that of Ref.~\cite{Mueller} except for a numerical
factor. 
Using the above parameters, 
we obtain $\lambda\sim18\mu m$, in reasonable
agreement with the observed value of about 40$\mu m$~\cite{Miesner}.

In conclusion, we have presented a versatile many-body theory of spin-1 
dilute bose gas. Depending on the parameters, a single effective 
Hamiltonian (\ref{Bog}) describes various many-body ground states, 
metastable spin domains, and the corresponding excitation spectra. 
An extension to spin-2 BEC can be similarly carried out and will be 
reported elsewhere.

This work was supported by a grant for Core Research for Evolutional
Science and Technology (CREST) of the Japan Science and Technology
Corporation (JST), by a Grant-in-Aid for Scientific Research 
(Grant No. 11216204) by the Ministry of Education, Science, Sports,
and Culture of Japan, and by the Toray Science Foundation.


\begin{thebibliography}{99}
\bibitem{Ueda} Detailed derivations will be published elsewhere.
\end{thebibliography}
\end{document}